\documentclass{emulateapj}

\usepackage{graphicx}
\usepackage{dcolumn}
\usepackage{bm}
\usepackage{epsfig}
\usepackage{subfigure}

\begin{document}

\topmargin 0.25in

\title{Fission Cycling in a Supernova $r$-process}

\author{J. Beun}
\affil{Department of Physics, North Carolina State University, Raleigh, NC 27695-8202}


\author{G. C. McLaughlin}
\affil{Department of Physics, North Carolina State University, Raleigh, NC 27695-8202}


\author{R. Surman}
\affil{Department of Physics, Union College, Schenectady, NY 12308}


\author{W. R. Hix}
\affil{Physics Division, Oak Ridge National Laboratory, Oak Ridge, TN 37831-6374}

\begin{abstract}

Recent halo star abundance observations exhibit an important feature of consequence to the $r$-process:  
the presence of a main $r$-process between the second and third peaks which is consistent among halo stars.
We explore fission cycling and steady-$\beta$ flow as the driving mechanisms behind this feature.
The presence of fission cycling during the $r$-process can account for nucleosynthesis yields between the second and third peaks, 
whereas the presence of steady-$\beta$ flow can account for consistent $r$-process patterns, 
robust under small variations in astrophysical conditions.
We employ the neutrino-driven wind of the core-collapse supernova to examine fission cycling and steady-$\beta$ flow in the $r$-process.
As the traditional neutrino-driven wind model does not produce the required very neutron-rich conditions for these mechanisms,
we examine changes to the neutrino physics necessary for fission cycling to occur in the neutrino-driven wind environment,
and we explore under what conditions steady-$\beta$ flow is obtained. 

\end{abstract}


\section{Introduction}
\label{Intro}

The production of neutron-capture elements early in the universe is recorded in the abundances of Galactic halo stars.
The two mechanisms responsible for generating most neutron-capture elements are the $r$-process and the $s$-process,
with the $r$-process generating about half of the nuclides with $A \gtrsim 100$ 
\citep{1957RvMP...29..547B,1957Cameron,1994ARA&A..32..153M}.
The observed $r$-process abundance distribution in metal-poor halo stars diverges 
above and below $Z \approx 58$ into two distinct patterns 
\citep{1995AJ....109.2757M,1996ApJ...467..819S,1998AJ....115.1640M,2000ApJ...544..302B,2000ApJ...530..783W,
2001Natur.409..691C,2002A&A...387..560H,2002ApJ...572..861C,
2003ApJ...591..936S,2004ApJ...607..474H,2004A&A...428.1027C,2007ApJ...660L.117F}.  
These patterns are defined by their consistency with the solar and halo star $r$-process abundances.
The low r-process pattern ($Z \lesssim 58$) is not consistent with either the solar $r$-process abundances
or among individual halo stars.
However, the high abundance ($58 \lesssim Z \lesssim 76$) pattern, known as the main $r$-process,
is very consistent among both the available halo star data and 
the solar system $r$-process abundances \citep{2006astro.ph.10412C,2007arXiv0705.4512A}.
We suggest the existence of a main $r$-process implies that a robust mechanism is producing elements from the second ($A \approx 130$) 
through the third ($A \approx 195$) $r$-process peak region \citep{Beun:2006vg,MartinezPinedo:2006fk}.

While there has been success in observing the abundance fingerprints of the $r$-process,
determining the astrophysical production site(s) remains an open question 
\citep{1989ApJ...343..254M,1992ApJ...399..656M,1994ApJ...433..229W,Thielemann:2001rn,2003MNRAS.345.1077R,2005NuPhA.758..587G}.
A promising candidate site for the $r$-process is the neutrino-driven wind of the post-bounce core-collapse supernova,
as the amount of $r$-process material ejected can account for the Galactic abundances \citep{1994ApJ...433..229W,2001ApJ...554..578W}.
Also, the frequency of the core-collapse supernova event is sufficiently rapid to produce $r$-process material
early in the universe, in accordance with the $r$-process abundances observed in very metal-poor stars \citep{argast-2004-416}.
However, traditional models of the neutrino-driven wind do not quite yield an $r$-process 
due to the low free neutron abundance at the $r$-process epoch \citep{1997ApJ...482..951H,2001ApJ...562..887T}.

The $r$-process requires a neutron-to-seed ratio of $R \gtrsim 100$ for production of the heaviest, $A \approx 195$,
peak \citep{1994ARA&A..32..153M}.
Uncertainties in the wind model may be responsible for this shortcoming as $R$ is related to both the entropy, $S$, 
and the dynamical timescale, $\tau$, and both are sensitive to the the proto-neutron star physics, 
see {\emph{e.g.}} \citet{2003PrPNP..50..153Q};
an increase of $S$ and a decrease of $\tau$ both lead to a higher initial neutron-to-seed ratio, $R$
\citep{1997ApJ...482..951H}.
These changes would likely require a compact and/or massive proto-neutron star
\citep{1996ApJ...471..331Q,2001ApJ...562..887T,1997ApJ...486L.111C,2000ApJ...533..424O,2001ApJ...554..578W}.
 Additional studies have looked to a prompt supernova explosion mechanism for producing the neutron-rich conditions necessary
for the $r$-process \citep{2001ApJ...562..880S,2003ApJ...593..968W},
although the physical realization of a prompt explosion scenario remains uncertain \citep{1989ApJ...340..955B,1989ApJ...341..385B}.

An avenue that has been suggested for generating a suitable neutron-to-seed ratio is
a shocked wind solution. A supersonic wind expansion, reducing seed formation during the $\alpha$-particle formation epoch,
is followed by a deceleration at late times \citep{Arcones:2006uq,2001ApJ...562..887T}
which allows the remaining free neutrons to be fully reincorporated into the $r$-process.
Recent $r$-process models in this supersonic outflow scenario can reproduce the solar system abundances,
but require an artificially high neutron-to-proton ratio \citep{2007arXiv0706.4360W}.
Alternate wind solutions generated solely by magnetic fields or acoustic waves,
{\emph e.g.} \citep{2007astro.ph..2539B,2007ApJ...659..561M},
also do not appear to produce a high enough neutron-to-seed ratio for an $r$-process.
Given these difficulties, alternative sites remain viable  
candidates, including neutron star mergers \citep{2003MNRAS.345.1077R} and gamma ray bursts \citep{2004ApJ...603..611S,Surman:2005kf}. 

In the neutrino-driven wind environment, charged-current neutrino reactions on both nucleons 
and nuclei determine the neutron-to-seed ratio for the $r$-process.
The relevant neutrino interactions on nucleons are
\begin{equation}\label{eq:nu_n}
\nu_e + {\rm n}  \rightleftharpoons {\rm p}+ e^{-} , 
\end{equation}
\begin{equation}\label{eq:anu_p} 
\bar{\nu}_e + {\rm p} \rightleftharpoons {\rm n} + e^{+} .
\end{equation}
For example, when electron anti-neutrino capture creating neutrons, Eqn. \ref{eq:nu_n}, 
is favored over electron neutrino capture creating protons,
Eqn. \ref{eq:anu_p}, a higher neutron-to-proton ratio results.
These reactions are sensitive to the hardness of the electron neutrino and anti-neutrino spectra,
which we parameterize by the effective electron neutrino and anti-neutrino temperature, $T_{\nu_e}$ and $T_{\bar{\nu}_e}$,
the electron neutrino and anti-neutrino luminosity, $L_{\nu_e}$ and $L_{\bar{\nu}_e}$,
and the distance from the center of the proto-neutron star. 
An analytic form of these neutrino-nucleon capture rates is found in \citet{1996ApJ...472..440M}.
Neutrino-nucleon reactions are responsible for the ``alpha'' effect,
a critical detriment to the $r$-process in the presence of a high electron neutrino flux
during the $\alpha$-particle formation epoch \citep{1995ApJ...453..792F,1996ApJ...472..440M,1998PhRvC..58.3696M}.
There have been several physical modifications proposed to reduce the efficacy of the ``alpha'' effect; 
for instance, a fast outflow wind \citep{1997ApJ...482..951H} or active-sterile neutrino oscillations 
\citep{1999PhRvC..59.2873M,2003APh....18..433F,2006PhRvD..73i3007B} lead to a successful $r$-process.
Both mechanisms reduce the capture of free neutrons by electron neutrinos during the $\alpha$-particle formation epoch,
allowing for a successful $r$-process.

When classifying the neutron richness of the $r$-process, we employ the electron fraction:
\begin{equation}\label{eq:yenp} 
Y_e = \frac{N_p}{N_p + N_n},
\end{equation}
where $N_n$ and $N_p$ are the total neutron and proton number densities
including both free nucleons and those in nuclei.
The neutron-to-proton ratio may be taken as $N_n / N_p$. 
Environments with a low electron fraction have the high free neutron densities 
necessary to drive the rapid capture of neutrons onto seed nuclei,
generating increasingly heavier nuclides by atomic weight, $A$.
During the $r$-process, these nuclei capture neutrons, become increasingly beta unstable,
and undergo $\beta$-decay to proceed to a larger atomic number, $Z$.
After $\beta$-decay, the nuclide can again capture neutrons and the cycle repeats.
As the supply of free neutrons dwindles, the nuclides decay back to $\beta$-stability.
The union of rapid neutron capture and $\beta$-decay results in an abundance pattern distinguished by 
large peak features at $A \approx 80,\ 130 \ {\rm and},\ 195$,
the first, second, and third peaks respectively.
These peaks correspond to the closed shell nuclei with slow $\beta$-decay rates,
creating ``waiting points'', causing material to accumulate at these points.

Abundances of the second and third $r$-process peaks, and the intermediate nuclei between them,
are defined by the nuclear properties of the $r$-process under conditions where both steady $\beta$-flow and fission cycling persist
\citep{1965ApJS...11..121S}.
Fission cycling occurs for sufficiently neutron-rich conditions where 
the $r$-process extends to nuclides that can decay through fission channels.
Fission impacts the $r$-process by terminating the path near the trans-uranium region.
This termination results in material returning to the $A \approx 130$ peak \citep{2005NuPhA.747..633P}.
If neutron-rich conditions persist, these fission products effectively become new seed nuclei for the $r$-process,
facilitating steady $\beta$-flow. 

Once a sufficient supply of both neutrons and seed nuclei are available,
steady-$\beta$ flow correlates the $r$-process abundances to their inverse $\beta$-decay rates,
see \emph{e.g.} \citet{1991PhR...208..267C} for further discussion.
The individual abundances in $N$ along an isotopic chain are determined by (n,$\gamma$) $\rightleftarrows$ ($\gamma$,n) equilibrium.
These correlated abundances span two or more of the closed-shell regions,
and when occurring between the second, $A \approx 130$, and third, $A \approx 195$, $r$-process peaks,
are a set of isotopic chains between $58 \lesssim Z \lesssim 76$.
The $Z$ of the last chain is determined by fission.
The first and last chains are linked as 
material leaving the last chain becomes the seed nuclides for the first chain through fission cycling.
The rate change of the total abundance of an isotopic chain, $Y(Z)$, 
is the difference between material entering and exiting the chain by $\beta$-decay,
\begin{eqnarray}\label{bflow1}
\dot{Y}(Z) & = & \sum_{A}^{} Y(Z-1,A)\lambda_{\beta}(Z-1,A) \\
& & {} -  \sum_{A}^{} Y(Z)\lambda_{\beta}(Z,A) \nonumber.
\end{eqnarray}
Above, the individual abundances of an isotopic chain are $Y(Z,A)$, 
and the individual $\beta$-decay rates are $\lambda_{_\beta}(Z,A)$.
If both (n,$\gamma$) $\rightleftarrows$ ($\gamma$,n) equilibrium and re-population of the seed nuclei persist,
the flow of material between isotopic chains reaches a steady-state configuration and 
the abundances and $\beta$-decay rates of each isotopic chain are inversely related as:
\begin{equation}\label{bflow2}
\sum_{A}^{} Y(Z,A)\lambda_{\beta}(Z,A) = {\rm const. }
\end{equation}
This is known as the steady $\beta$-flow condition.

Here we examine fission cycling and steady-$\beta$ flow mechanisms 
and discuss their impact on the $r$-process in the neutrino-driven wind.
We also explore changes to the neutrino physics, independent of a physical generator,
that lead to environments where fission cycling and steady-$\beta$ flow mechanisms are present in the neutrino-driven wind.
In section \ref{nwind}, we describe the details of our calculation in the neutrino-driven wind.
In section \ref{Fission_Cycling} we describe how fission cycling influences the $r$-process.   
Section \ref{steadyflow} describes how steady-$\beta$ flow leads to a robust $r$-process pattern.
In section \ref{nchanges} we detail how the neutrino spectrum influences the $r$-process environment.
In section \ref{conclusions}, we summarize the results.

\section{Description of Nucleosynthesis Modeling}
\label{nwind}

The neutrino-driven wind forms several seconds post-core bounce in the core-collapse supernova environment,
and a one-dimensional wind model is often employed to describe the abundance composition of the isotropic outflow
\citep{1994ApJ...433..229W,1986ApJ...309..141D}.
Two-dimensional wind models have been employed previously;
however, these calculations do not self-consistently  
account for neutrino interactions, necessitating the use of an artificial $Y_e$ \citep{2006ApJ...646L.131F}.
Here we follow the same wind parameterization as \citet{2006PhRvD..73i3007B}.
Unless otherwise noted, our calculations use an entropy per baryon of $s/k = 100$, 
an outflow timescale of $\tau = 0.3 \, \ {\rm s}$,
an initial density of $\rho_o \approx 1.7 \times 10^8 \ {\rm g/cm^3}$,
and an initial radius of $r_7 \approx 0.1$, in units of $10^7 \ {\rm cm}$.

We track the abundance composition of a mass element, following ejection from the proton-neutron star,
using three coupled reaction networks.
The element is initially in Nuclear Statistical Equilibrium (NSE),
and we follow it using a NSE network \citep{1999PhRvC..59.2873M} until $\alpha$-particles begin to form, $T_9 \approx 10$,
where $T_9$ is in units of $10^9$ K.
Next, we track the mass element throughout $\alpha$-particle formation with an intermediate network calculation \citep{hix1999} 
that includes strong and electromagnetic rates tabulated by \citet{2000ADNDT..75....1R}. 
We have added electron-capture, positron-capture, neutrino-capture, and anti-neutrino-capture rates from \citet{1995ApJ...455..202M}
to both networks.
As material reaches the $r$-process epoch, $T_9 \approx 2.5$,
we use an $r$-process network \citep{2001PhRvC..64c5801S,PhysRevLett.79.1809}
that handles the relevant reactions of $\beta$-decay, $\beta$-delayed neutron emission, neutron capture, and photo-disintegration, 
and we include charged-current neutrino interactions.
We use $\beta$-decay rates and neutron separations from \citet{1997ADNDT..66..131M}, 
neutron capture rates from \citet{2000ADNDT..75....1R}, and neutrino-capture rates from \citet{1995ApJ...455..202M}.
A complete description of the network calculation can be found in \citet{2006PhRvD..73i3007B}.

\section{Fission Cycling}
\label{Fission_Cycling}
Fission decay joins $\beta$-decay, neutron capture, and photo-dissociation as a primary reaction channel
for $r$-process nuclides near the trans-uranium region, as previously mentioned.
Nuclei that undergo fission provide a termination point for the $r$-process 
by preventing the production of heavier nuclides past this region.
After the nuclide undergoes fission, 
it fragments into smaller nuclides that rejoin the $r$-process in the $A \approx 130$ region.
This leads to fission cycling, with the fission daughter products acting as seed nuclei for the $r$-process.

A knowledge of both the location of fission and the resultant distribution of fission fragments
are necessary for detailed calculations of the $r$-process in very neutron-rich environments.
Treatments of fission in the $r$-process 
\citep{1987ApJ...323..543C,1994ApJ...429..499R,1999ApJ...521..194C,1999ApJ...516..381F} 
have been limited to largely phenomenological methods 
since there are few measurements of the fission properties for neutron-rich heavy nuclides. 
This current study is concerned with the general effects of fission on the $r$-process.
We employ $\beta$-delayed fission probabilities from \citet{hilf76}
which are relevant to our mass model and astrophysical conditions.
We indirectly include the effects of neutron-induced fission, 
as no complete and consistent set of neutron-induced rates is available,
by employing a spontaneous fission region for $A \gtrsim 270$, where all of the nuclides instantaneously fission,
following previous treatments described in \citet{2004AstL...30..647P}.
We implement fission by depopulating the parent fission nuclide into its appropriate daughter nuclides at
the end of each timestep in our network code.
The effects of neutrino-induced fission are small and are not included \citep{2004PhRvL..92k1101K,2004ApJ...608..470T}. 
We employ phenomenological fission distributions,
following previous studies described in \citet{2004AstL...30..647P},
to elucidate the primary effects fission cycling has on the $r$-process pattern.
We also include fission-induced neutron emission when noted,
implementing a piecewise linear fit of the values from \citet{2005PhLB..616...48K}.
We treat the emitted neutrons as becoming immediately thermalized with the surrounding medium.   

As the fission material is re-incorporated in the $A \approx 130$ peak,
this material experiences ``waiting points'' and accumulates here as 
$\beta$-decay is necessary for material to move from one isotopic ($Z$) chain to the next ($Z + 1$) chain. 
The material is initially compacted together in atomic mass space, $A$, and then disperses through $\beta$-decay.
Note that some material remains in each peak region during the fission cycling process
as $\beta$-decay does not fully deplete each peak.

\begin{figure}
\epsscale{1.25}
\plotone{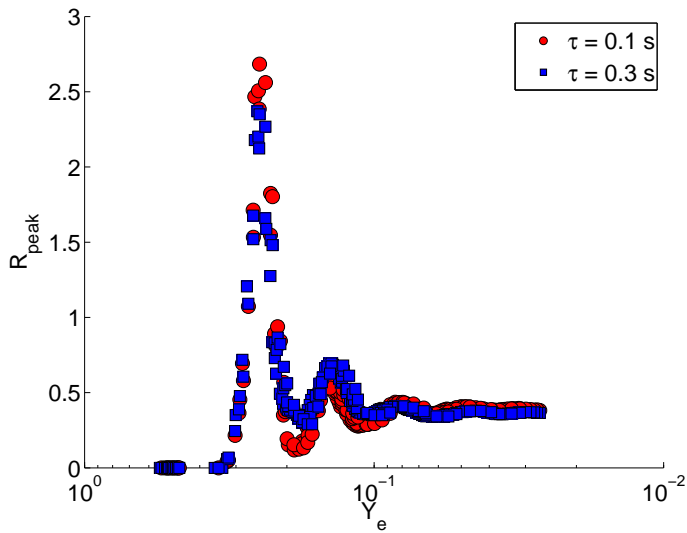} 
\caption{The final $R_{{\rm peak}}$ (Eqn. \ref{eq:rpeak}) resulting from a symmetric fission distribution 
for two different outflow timescales, $\tau = 0.1 {\rm \ s}$ (circles) and $\tau = 0.3 {\rm \ s}$ (squares),
is shown for a variety of electron fractions, $Y_e$.
For very neutron-rich conditions, $Y_e \lesssim 0.1$, 
a consistent $r$-process pattern forms between the second, $A \approx 130$, and third, $A \approx 195$, peak region.
Fission cycling during the $r$-process links the second and third peaks,
as material that captures out of the third peak reaches the fission regime.  
The resulting fission daughter products then rejoin the $r$-process at the second peak.
\label{ynpeak}}
\end{figure}

To examine the accumulation of material during this $\beta$-decay dispersion process,
we compare abundances in the second and third $r$-process peaks,
\begin{equation}\label{eq:rpeak} 
R_{{\rm peak}} = \frac{\displaystyle\sum_{A=190}^{200} Y(A) }{ \displaystyle\sum_{A=125}^{135} Y(A)}.
\end{equation}
Figure \ref{ynpeak} demonstrates $R_{{\rm peak}}$ for a range of $Y_e$.
The $Y_e$ is from the start of of the $r$-process epoch.
Each point represents an individual nucleosynthesis outcome resulting from a unique choice of electron 
neutrino and anti-neutrino luminosities in the neutrino-driven wind.
The left-most portion ($Y_e \gtrsim 0.3$) of Fig. \ref{ynpeak}, where $R_{{\rm peak}}$ is zero, 
denotes where final $r$-process abundances extend from the second peak and up to the rare earth region.
The first bump, at $Y_e \approx 0.3$, 
denotes patterns where the heaviest $r$-process nuclides occupy the third peak and have partially left the second peak.
Once past this bump, the heaviest $r$-process nuclides are beyond the third peak and have reached the fission region,
and re-incorporate into the second peak. 

The behavior of fission under very neutron-rich conditions can be quantified by the number of fission cycles.
When a nuclide fissions, there is an increase of heavy nuclei in abundance as multiple daughter nuclides are produced.
The doubling of the total abundance is a fission cycle,
\begin{equation}\label{eq:fiscycle} 
\Lambda = \log_2(Y_{end}) - \log_2(Y_{start}),
\end{equation}
where $\Lambda$ is the number of fission cycles and $Y_{start}$ and $Y_{end}$ are the total abundance before and after 
fission cycling respectively.
In Fig. \ref{ynpeak}, cycles occur at the minima past each bump and extend to the next minima.
For example, the first fission cycle ranges from the minima at $Y_e \approx 0.17$ to the minima at $Y_e \approx 0.09$.
Significant fission cycling occurs for increasingly neutron-rich conditions, and, under very neutron-rich conditions,
the movement of material entering and leaving a peak reaches equilibrium, a consequence of steady $\beta$-flow.

The equilibrium $R_{{\rm peak}}$, due to fission cycling, 
appears as the straight region of the curve in Fig. \ref{ynpeak} for $Y_e \lesssim 0.1$,
and does not change significantly for variations in the wind conditions.
This is depicted in Fig. \ref{ynpeak}
as the equilibrium peak ratio is not strongly affected by choice of the outflow timescale.
Here, this stability is demonstrated for the wind outflow timescales of  $\tau = 0.1 {\rm \ s}$ and $\tau = 0.3 {\rm \ s}$.

The equilibrium value of $R_{{\rm peak}}$ is sensitive to properties of the fission model.
In particular,
it is sensitive to the specific location within the second peak to which fission returns material,
as the presence of waiting points here dictate the flow and accumulation of material.
Material deposited at or below closed-shell nuclei flow to a waiting point,
and remain in this peak region longer than material deposited above the closed-shell nuclei.
The accumulation of material in the $A \approx 130$ peak leads to smaller values of $R_{{\rm peak}}$.
Material arriving above the closed-shell bypasses the waiting point and continues flowing to heavier regions,
leading to larger values of $R_{{\rm peak}}$.

\begin{figure}
\epsscale{1.25}
\plotone{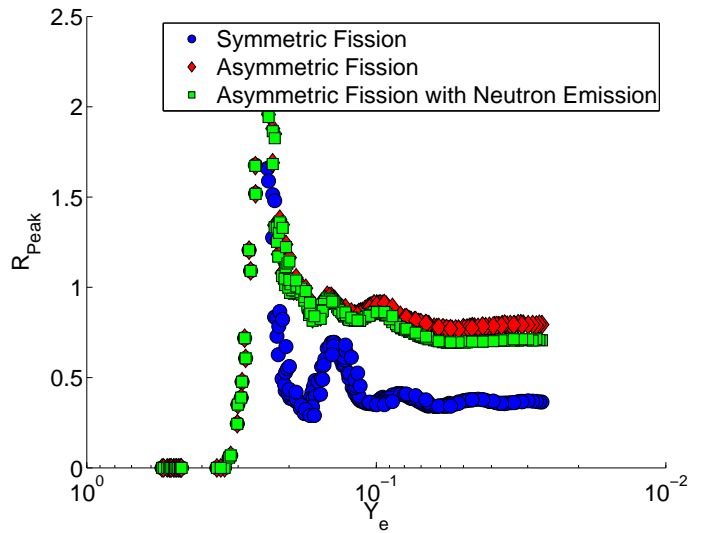}
\caption{Same as Fig. \ref{ynpeak},
but compares the effects of different fission daughter product distributions.  
The distribution of daughter products determines if material is deposited 
above or below the closed-shell nuclei in the $A \approx 130$ peak.
Fission distributions depositing material above the closed-shell nuclides, asymmetric fission (diamonds),
leads to a higher equilibrium $R_{{\rm peak}}$, 
as more material is cycled through the $A \approx 130$ peak.
Distributions depositing material below the peak, symmetric fission (circles),
have a lower equilibrium $R_{{\rm peak}}$ and cycle less material through the peak region.
Fission-induced neutron emission deposits additional material below the closed-shell nuclides, lowering $R_{{\rm peak}}$,
and is shown above for asymmetric fission (squares).
\label{peakyefis}}
\end{figure}

In Fig. \ref{peakyefis} we show specific examples of the 
consequences to $R_{{\rm peak}}$ resulting from different fission daughter product distributions.
We consider cases of fission daughter product distributions with both symmetric and asymmetric modes.
Symmetric distributions have daughter nuclides that are close in both charge, $Z$, and mass, $A$.
These distributions deposit more material at or below the closed-shell nuclides
and have a smaller equilibrium $R_{{\rm peak}}$, shown as circles in Fig. \ref{peakyefis}.
Asymmetric distributions have daughter product distributions whose charge and mass are separated
proportionally by a scaling factor, resulting in one daughter being larger in both $A$ and $Z$ than the other.
This leads to more material being deposited above the closed-shell nuclides
and a higher equilibrium $R_{{\rm peak}}$, shown as diamonds in Fig. \ref{peakyefis}.
When neutrons are emitted during the course of fission, fewer daughter products lie above the close-shell nuclei,
lowering the equilibrium $R_{{\rm peak}}$ from the asymmetric case, shown as squares in Fig. \ref{peakyefis}.
Detailed knowledge of both the fission location and daughter product distributions 
is required to fully model the $r$-process abundances in very neutron-rich environments. 

\begin{figure}
\epsscale{1.25}
\plotone{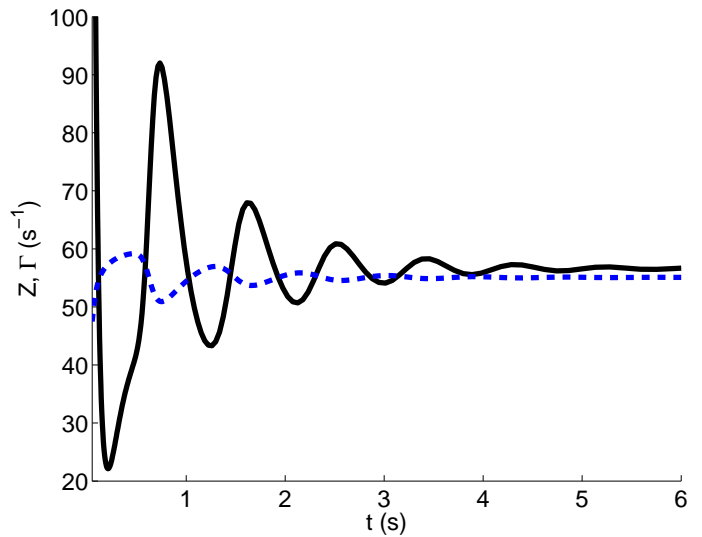}
\caption{ Under very neutron-rich conditions the effective decay rate of the peaks oscillate until  
equilibrating at the steady $\beta$-flow rate.
The abundance weighted atomic number, $Z = \displaystyle\sum_{Z=45}^{60} Z \ Y(Z) / \displaystyle\sum_{Z=45}^{60} Y(Z)$,
 (dashes) is plotted with the abundance weighted $\beta$-decay rate, 
$\Gamma = \displaystyle\sum_{Z=45}^{60} \Gamma_{\beta}(Z) \ Y(Z) / \displaystyle\sum_{Z=45}^{60} Y(Z)$, 
of an isotopic chain (solid) for the second, $A \approx 130$, peak region,
versus time, t.
The oscillation of the decay rates in the peak regions are due to 
the changing population of different nuclides during the course of fission cycling.
To elucidate abundance changes between isotopic chains, the data above results from our phenomenological model,
Eqn. \ref{eq:decay_model}, under conditions with an $Y_e = 0.05$ at the start of the $r$-process epoch ($T_9 \approx 2.5$).
\label{fullbeta}}
\end{figure}

The change in abundance of the various $r$-process peaks during fission cycling, as depicted in Fig. \ref{ynpeak} and \ref{peakyefis}, 
can be phenomenologically described by the effective rate of material entering and exiting the major peak regions.
The effect of fission cycling on the peak abundances is modeled
by approximating the flow of fission material as rapid compared to the flow leaving the $A \approx 130$, the rare earth, 
and the $A \approx 195$ peak regions.  
The rate of abundance change in each peak becomes:
\begin{eqnarray}
\label{eq:decay_model}
\dot Y_{130} & = & - \Gamma_{130}(t) Y_{130} + f \Gamma_{195}(t) Y_{195} \\  
\dot Y_{{\rm Earth}} & = & - \Gamma_{{\rm Earth}}(t) Y_{{\rm Earth}} + \Gamma_{130}(t) Y_{130} \\
& & {} + (2 - f) \Gamma_{195}(t) Y_{195} \nonumber \\  
\dot Y_{195} & = & - \Gamma_{195}(t) Y_{195} + \Gamma_{{\rm Earth}}(t) Y_{{\rm Earth}}
\end{eqnarray}
where $Y$ is the abundance of each peak and $\Gamma$ is the rate of decay leaving a region,
with the subscripts corresponding to the $A \approx 130$, rare earth, and $A \approx 195$ peaks.
Here $f$ is the distribution of fission daughter products between the $A \approx 130$ and rare earth regions.
For purposes of illustrating the toy model, we take all fission products as arriving to the $A \approx 130$ region, $f=2$.
The equilibrium (steady $\beta$-flow) behavior of fission cycling is depicted in the far right-hand region of Fig. \ref{ynpeak}.
For discussion of solutions for long-time fission cycling see \citet{1965ApJS...11..121S}.
The transient region prior to steady $\beta$-flow equilibrium is dependent on the flow (decay rate)
of material leaving each peak region, $\Gamma_{130}$, $\Gamma_{{\rm Earth}}$, and $\Gamma_{195}$.
The effective flow out of a region is determined by the $\beta$-decay rate of each isotopic chain,
as material is in $(n,\gamma) \leftrightharpoons (\gamma,n)$ equilibrium during this fission cycling phase.
Additionally, this outflow is determined by the population of nuclides within a peak,
since the individual $\beta$-decay rate changes from nuclide to nuclide.
As a consequence, various nuclides are populated while material is flowing through each peak,
leading to changes in the effective flow out of each peak.
For example, the effective flow rate in the $A \approx 130$ peak slows 
as the closed shell nuclides, with slow $\beta$-decay rates, are populated.
We employ our phenomenological model to highlight the change in the effective flow rate in Fig. \ref{fullbeta},
for the $A \approx 130$ peak region with a $Y_e = 0.05$ at the start of the $r$-process.
Here, the fluctuation in the decay rates of a peak corresponds to the changing of individual nuclide abundances.
The movement of material within a peak is demonstrated by the corresponding change of the weighted atomic number, $Z$.
As the conditions in Fig. \ref{fullbeta} are very neutron-rich, sufficient for steady $\beta$-flow,
the right-most portion depicts a straight line for both the effective flow rate and $Z$,
since steady $\beta$-flow leads to isotopic chain abundances determined by the $\beta$-decay rate of each chain.

\section{Steady Beta Flow}
\label{steadyflow}

\begin{figure}
\epsscale{1.25}
\plotone{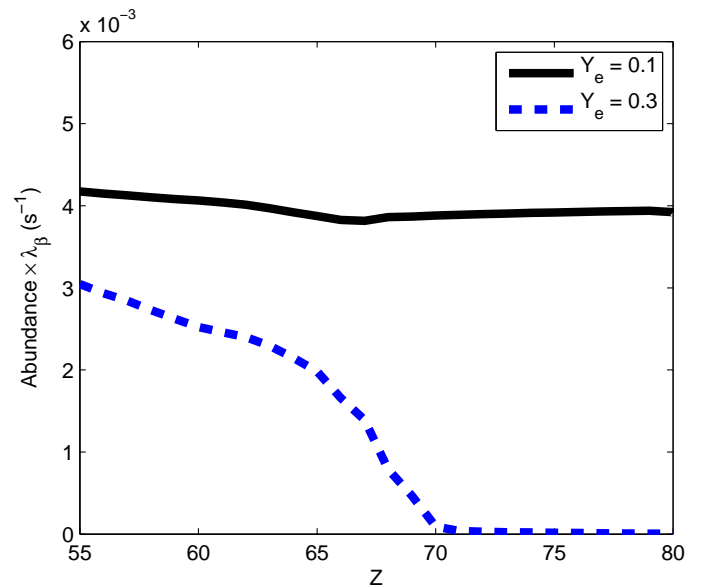}
\caption{We plot the steady $\beta$-flow condition, $\displaystyle\sum_{A}^{} Y(Z,A)\lambda_{\beta}(Z,A)$, 
versus atomic number, $Z$, for two different $Y_e$'s in the neutrino-driven wind.
For the case with an $Y_e = 0.1$ at the start of the $r$-process epoch,
conditions are sufficiently neutron-rich for steady $\beta$-flow,
marked by a straight line.
The case with $Y_e = 0.3$ is not neutron-rich enough for steady $\beta$-flow to obtain.
\label{pathplot}}
\end{figure}

As discussed in Sec. \ref{Intro}, the abundances of the $r$-process become fully determined under conditions where both
fission cycling and steady $\beta$-flow occur.
We now examine consequences to the final $r$-process abundances resulting from steady $\beta$-flow.
In Fig. \ref{pathplot}, the steady-$\beta$ flow condition, Eqn. \ref{bflow2},
is tested and fulfilled for very neutron-rich conditions in the neutrino-driven wind, 
appearing as a straight line between the second and third peaks of the $r$-process, the main $r$-process region.
For less neutron-rich conditions, steady-$\beta$ flow does not obtain as there are not enough free neutrons to sustain 
re-population of the seed nuclei by fission cycling. 

\begin{figure}
\epsscale{1.25}
\plotone{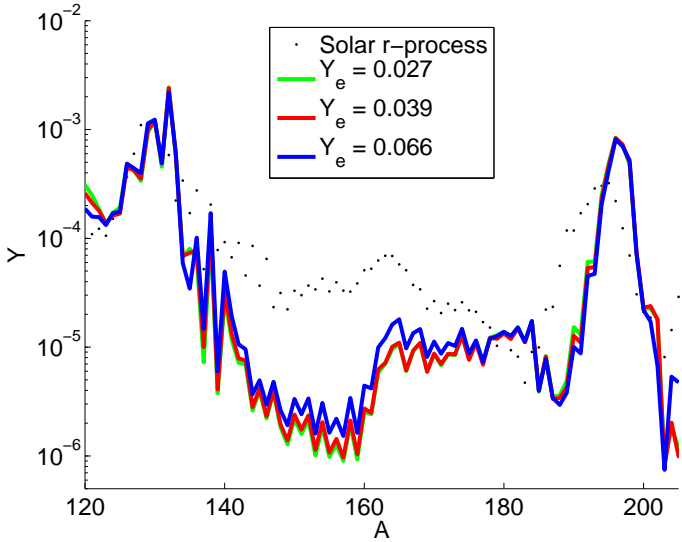}
\caption{When conditions neutron-rich enough for steady $\beta$-flow occur, 
a consistent $r$-process pattern emerges regardless of the initial $Y_e$.
The details of the abundance pattern are dependent on the nuclear physics employed in the mass model, as discussed in the text.
The abundance, $Y$, is plotted versus the atomic number, $Z$, for these very neutron-rich conditions.
The blue line is generated in the neutrino-driven wind for $L_{\nu_e} = 0.02$ and $L_{\overline{\nu}_e} = 3.0$,
the red line for $L_{\nu_e} = 0.01$ and $L_{\overline{\nu}_e} = 4.0$, and
the green line for $L_{\nu_e} = 0.006$ and $L_{\overline{\nu}_e} = 6.0$.
The electron neutrino and anti-neutrino luminosities, $L_{\nu_e}$ and $L_{\overline{\nu}_e}$, 
are in units of $\times 10^{51} \ {\rm \ ergs \ s^{-1}}$.
The effective temperature of the electron neutrinos is $T_{\nu_e} = 3.5 \ {\rm MeV}$ and for the electron anti-neutrinos is
$T_{\overline{\nu}_e} = 4.5 \ {\rm MeV}$.
Labeled is $Y_e$ at the start of the $r$-process epoch.
The black line represents the solar system abundances from \citet{Simmerer:2004jq},
and the gray diamonds are the $r$-process abundances from the halo star HD221170 \citep{2006ApJ...645..613I}.
All abundances are scaled to $10^{-4}$ at $Z=52$.
\label{abundance_plot}}
\end{figure}

A set of typical abundances resulting from steady-$\beta$ flow, 
occurring for very neutron-rich conditions in the neutrino-driven wind, are depicted in Fig. \ref{abundance_plot}.
The individual $Y_e$'s are produced by a unique choice of initial neutrino and anti-neutrino luminosities,
in the same manner as Fig. \ref{ynpeak}.
Our abundance patterns reproduce the general peak structure of the main $r$-process between the second and third peaks 
and is robust over a wide range of conditions in the neutrino-driven wind, 
directly resulting from the pairing of both fission cycling and steady $\beta$-flow.
Improvement to a nuclide by nuclide abundance comparison 
between calculation and both solar and halo star abundance data requires further understanding of 
the nuclear properties of nuclei far from stability
\citep{1996PhLB..387..455P,PhysRevC.52.R23}.
Additionally, our comparison may be affected by post-processing of the material by neutrinos,
as well as contributions from a weak or secondary $r$-process.
However, fission cycling in the neutrino-driven wind remains a viable explanation 
for the global properties of the observed main $r$-process
and warrants further investigation. 

\section{Influence of Neutrinos on the Electron Fraction}
\label{nchanges}

After neutrinos extricate material from the surface of the proto-neutron star,
the newly released material is composed entirely of free nucleons.
Although we dynamically calculate the $Y_e$ in our models,
it is pedagogically useful to examine the electron fraction through the approximation of weak equilibrium.
Since neutrino and anti-neutrino rates dominate electron and positron capture,
the initial weak equilibrium $Y_e^{(0)}$ is
\begin{equation}
\label{eq:ye_neutrino}
Y_e^{(0)} \approx \frac{1}{1 + \lambda_{\bar{\nu}_e} /
\lambda_{\nu_e}}.
\end{equation}

The rates of electron neutrino capture on neutrons and electron anti-neutrino capture on protons are $\lambda_{\nu_e}$ and
$\lambda_{\overline{\nu}_e}$ respectively.
An increase in the electron anti-neutrino capture rate, achievable by either  
boosting the electron anti-neutrino flux or hardening the electron anti-neutrino  
spectra, will tend to decrease Ye, while an increase in the neutrino  
capture rate will tend to increase Ye. 

Farther away from the proto-neutron star, as material in the expansion continues to cool,
$\alpha$-particles form and contribute to the weak equilibrium $Y_e$ approximation,
\begin{equation}
\label{eq:ye_alpha}
Y_e \approx Y_e^{(0)} + \left( \frac{1}{2} - Y_e^{(0)} \right)
X_{\alpha} .
\end{equation}
Here, $X_{\alpha}$ represents the mass fraction of $\alpha$-particles.

The formation of $\alpha$-particles is enhanced in the presence of a strong electron neutrino flux, 
as electron neutrinos capturing on free neutrons create new effective proton seeds for $\alpha$-particle formation.
These effective seeds quickly capture additional neutrons which continues to drive the $Y_e$ to $1 / 2$.
The depletion of neutrons from both neutrino capture and $\alpha$-particle formation is known as the alpha effect, and
this alpha effect must be removed for neutron-rich conditions to obtain for the $r$-process.

We look to neutrino luminosities and spectral properties for the removal of the alpha effect.
Since weak equilibrium does not fully obtain in the neutrino-driven wind,
we use the following procedure to calculate $Y_e$ instead of the weak equilibrium approximation described above.
Starting from a prospective set of neutrino effective temperatures and luminosities,
we adjust the initial Fermi-Dirac neutrino spectrum.
Then we self-consistently account for both the evolution of the updated neutrino spectrum in the neutrino-driven wind
and the relevant changes to the abundance.
From this, we dynamically account for the $Y_e$ throughout the nucleosynthesis epochs.
For a full description see \citet{2006PhRvD..73i3007B}.

\begin{figure}
\centering
\subfigure[] 
{
    \label{fig:yecontoura}
    \includegraphics[width=8cm]{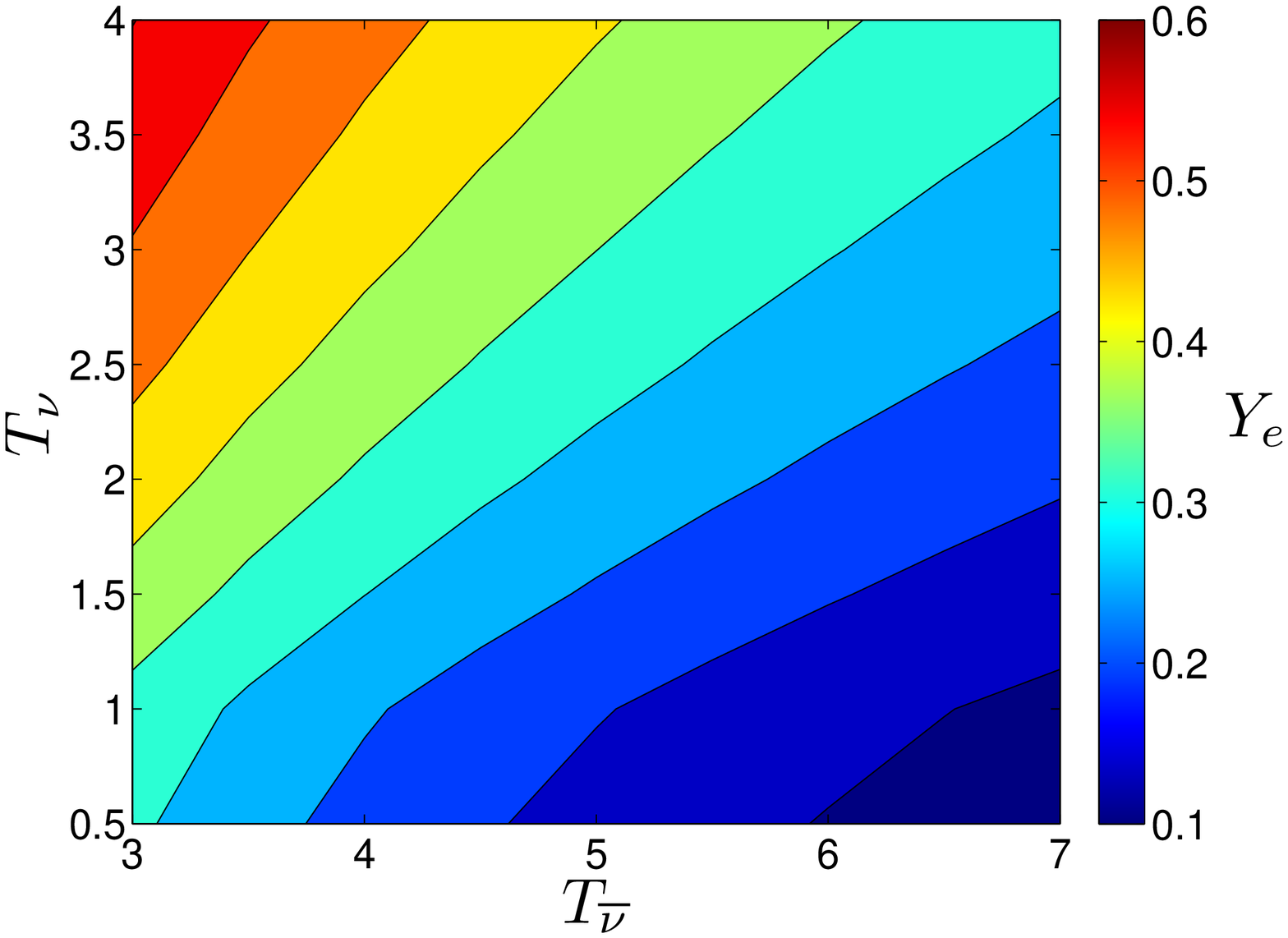}
}
\hspace{1cm}
\subfigure[] 
{
    \label{fig:yecontourb}
    \includegraphics[width=8cm]{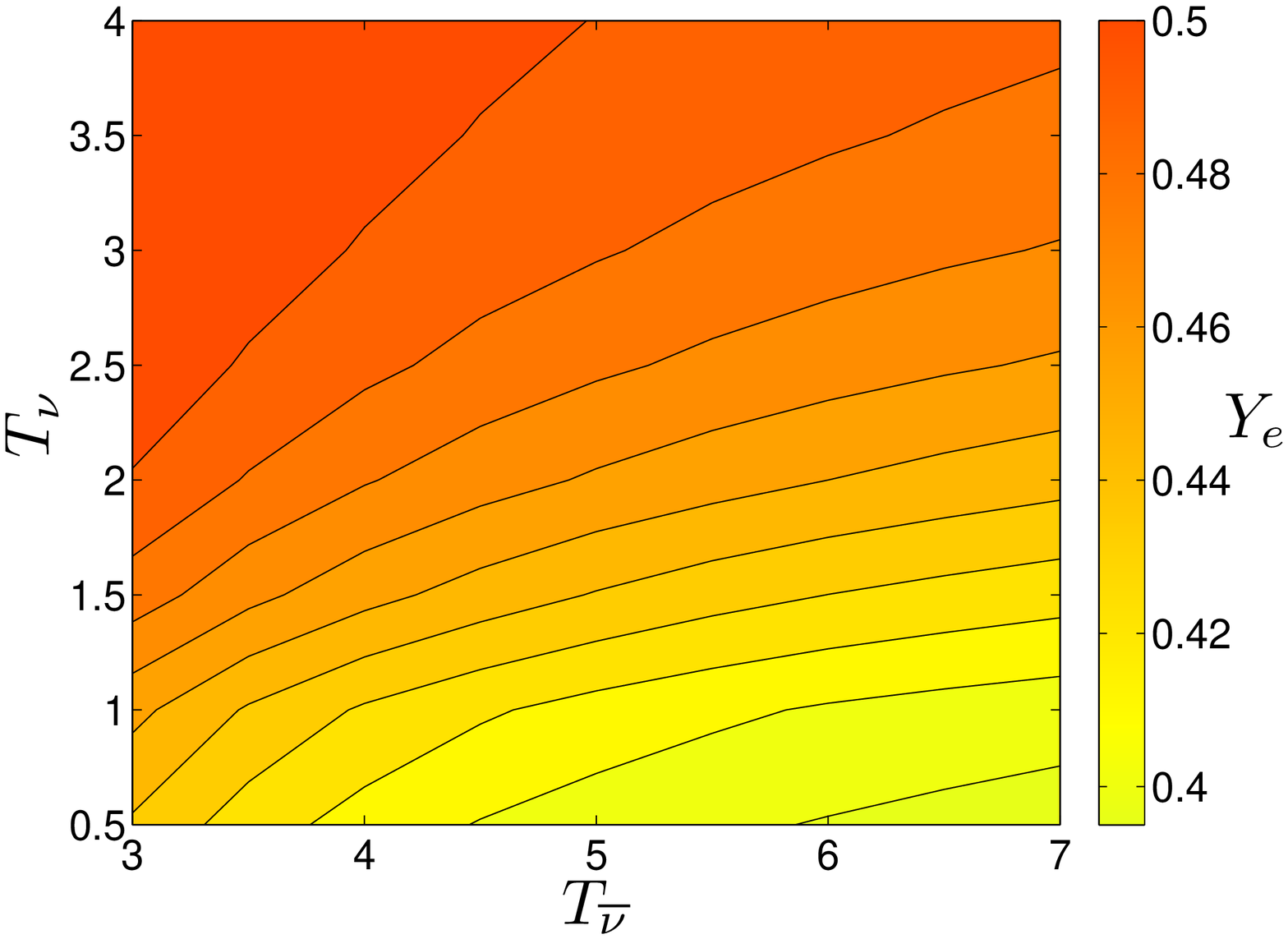}
}
\caption{The electron fraction, $Y_e$, is shown over a range of electron neutrino and anti-neutrino temperatures,
$T_{\nu}$ and $T_{\overline{\nu}}$. 
The ``alpha'' effect equally binds protons and neutrons into $\alpha$-particles 
which drives the electron fraction to $Y_e \approx 1/2$ and prevents an $r$-process.
Regions of $Y_e$ before $\alpha$-particle formation ($T_9 \approx 9.4$), Fig. \ref{fig:yecontoura}, 
initially appear favorable to the $r$-process; 
however, the ``alpha'' effect has pushed $Y_e$ near $1/2$ by the start of the $r$-process epoch ($T_9 \approx 2.5$),
Fig. \ref{fig:yecontourb}. 
The effective electron neutrino and anti-neutrino temperatures, $T_{\nu}$ and $T_{\overline{\nu}}$, are in units of MeV.
The electron neutrino luminosity is $L_{\nu} = 1 \times 10^{51} \ {\rm \ ergs \ s^{-1}}$ and the electron anti-neutrino luminosity
is $L_{\overline{\nu}} = 1.3 \times 10^{51} \ {\rm \ ergs \ s^{-1}}$.}
\label{fig:yecontours} 
\end{figure}

We first examine $Y_e$'s resulting from a range of neutrino effective temperatures. 
Before the formation of $\alpha$-particles, $T_9 \approx 9.4$,
the $Y_e$'s from our model are consistent with weak equilibrium.
The $Y_e$ increases with larger electron neutrino effective temperatures as
these higher temperatures result in more protons being converted to neutrons.
This results from electron neutrino capture on neutrons, Eqn. \ref{eq:nu_n}.
Conversely, larger anti-neutrino energies produce a lower $Y_e$, 
a consequence of electron anti-neutrino capture, Eqn. \ref{eq:anu_p}.
The ranges of neutrino effective temperatures depicted in Fig. \ref{fig:yecontoura} 
result in $Y_e$ between $0.1 \lesssim Y_e \lesssim 0.6$, 
and the low $Y_e$ portion of the region seems initially promising for a successful $r$-process.
In Fig. \ref{fig:yecontours}, the luminosity of the electron neutrinos is $1 \times 10^{51} {\rm \ ergs \ s^{-1}}$ and
the electron anti-neutrino luminosity is $1.3 \times 10^{51} {\rm \ ergs \ s^{-1}}$.

After $\alpha$-particle formation and charged particle reactions have occurred, conditions generated by changes to the 
neutrino effective temperatures do not lead to a successful $r$-process.
The $Y_e$ at the $r$-process epoch is shown in Fig. \ref{fig:yecontourb},
and is noticeably compressed by the alpha effect, ranging between $0.4 \lesssim Y_e \lesssim 0.5$.
The $Y_e$ is increased compared to the previous epoch as
the alpha effect has converted free neutrons into additional effective proton seed nuclei.
These values of $Y_e$ are not neutron-rich enough to lead to a successful $r$-process.

The second case we examine is for changes in the neutrino and anti-neutrino luminosities.
Although initially similar, they lead to a decidedly different outcome,
as environments where sufficient reductions in the electron neutrino luminosity occur can lead to a successful $r$-process.
When both the neutrino and anti-neutrino luminosities are significantly reduced compared to traditional models,
electron capture and positron capture become important and influence the electron fraction,
\begin{equation}
\label{eq:yeeqel}
Y_e \approx 1/[1 + (\lambda_{\bar{\nu}_e}+\lambda_{e^-})
/(\lambda_{\nu_e} + \lambda_{e^+} )],
\end{equation}
where $\lambda_{e^-}$ and $\lambda_{e^+}$ are the electron and positron rates respectively \citep{1996ApJ...472..440M}.
When electron-positron pairs are important, as they are here, these rates tend to increase the $Y_e$
and begin influencing the $Y_e$ when the neutrino luminosities are
$L_{\nu_e} \lesssim 10^{50} {\rm \ ergs \ s^{-1}}$ and 
$L_{\overline{\nu}_e} \lesssim 5 \times 10^{50} {\rm \ ergs \ s^{-1}}$.
For wind solutions where neutrinos are less prevalent,
a weak equilibrium $Y_e$ dictated by electron and positron capture would not be sufficiently neutron-rich
for the $r$-process. 

\begin{figure}
\centering
\subfigure[] 
{
    \label{fig:lumyea}
    \includegraphics[width=8cm]{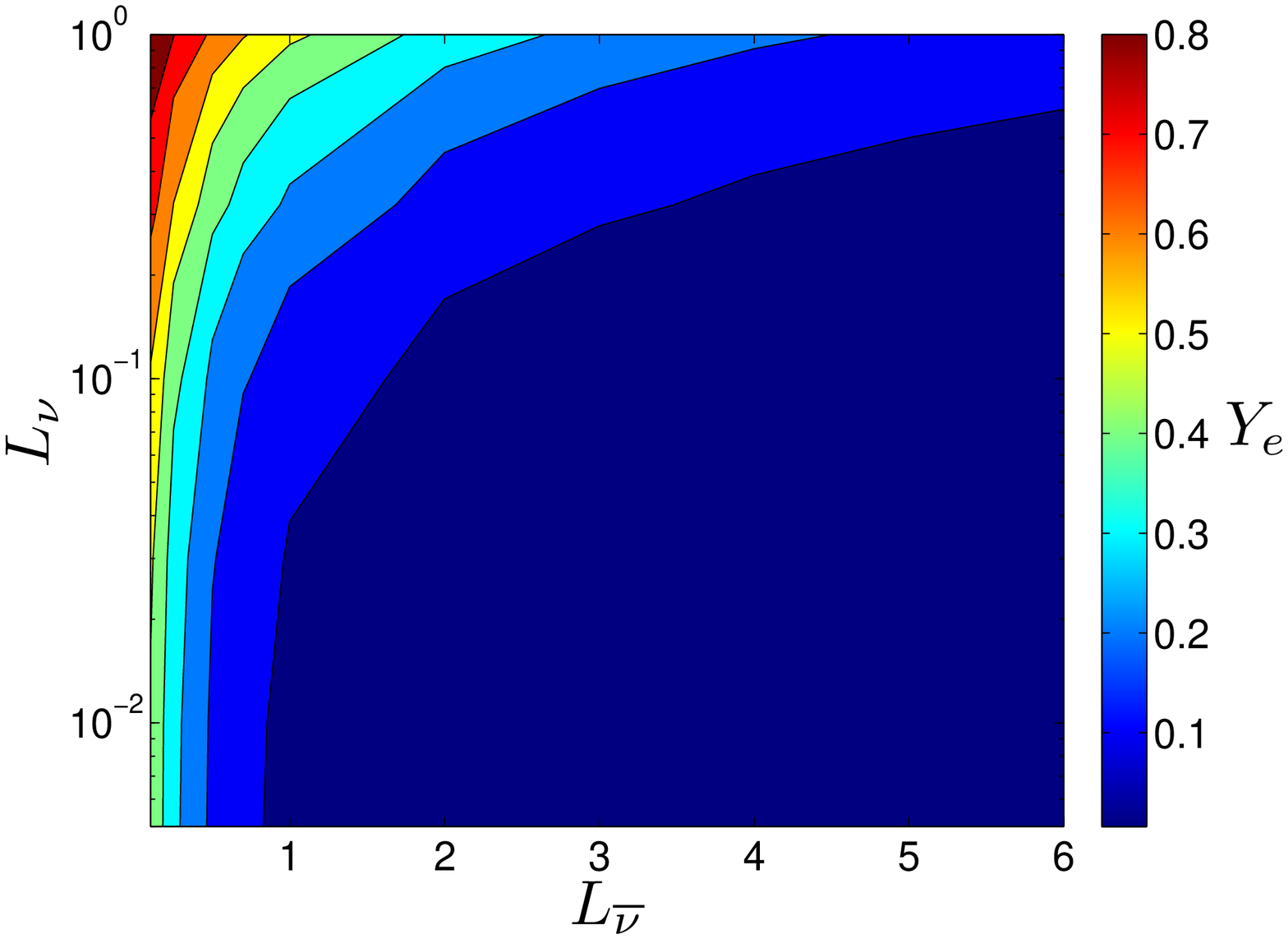}
}
\hspace{1cm}
\subfigure[] 
{
    \label{fig:lumyeb}
    \includegraphics[width=8cm]{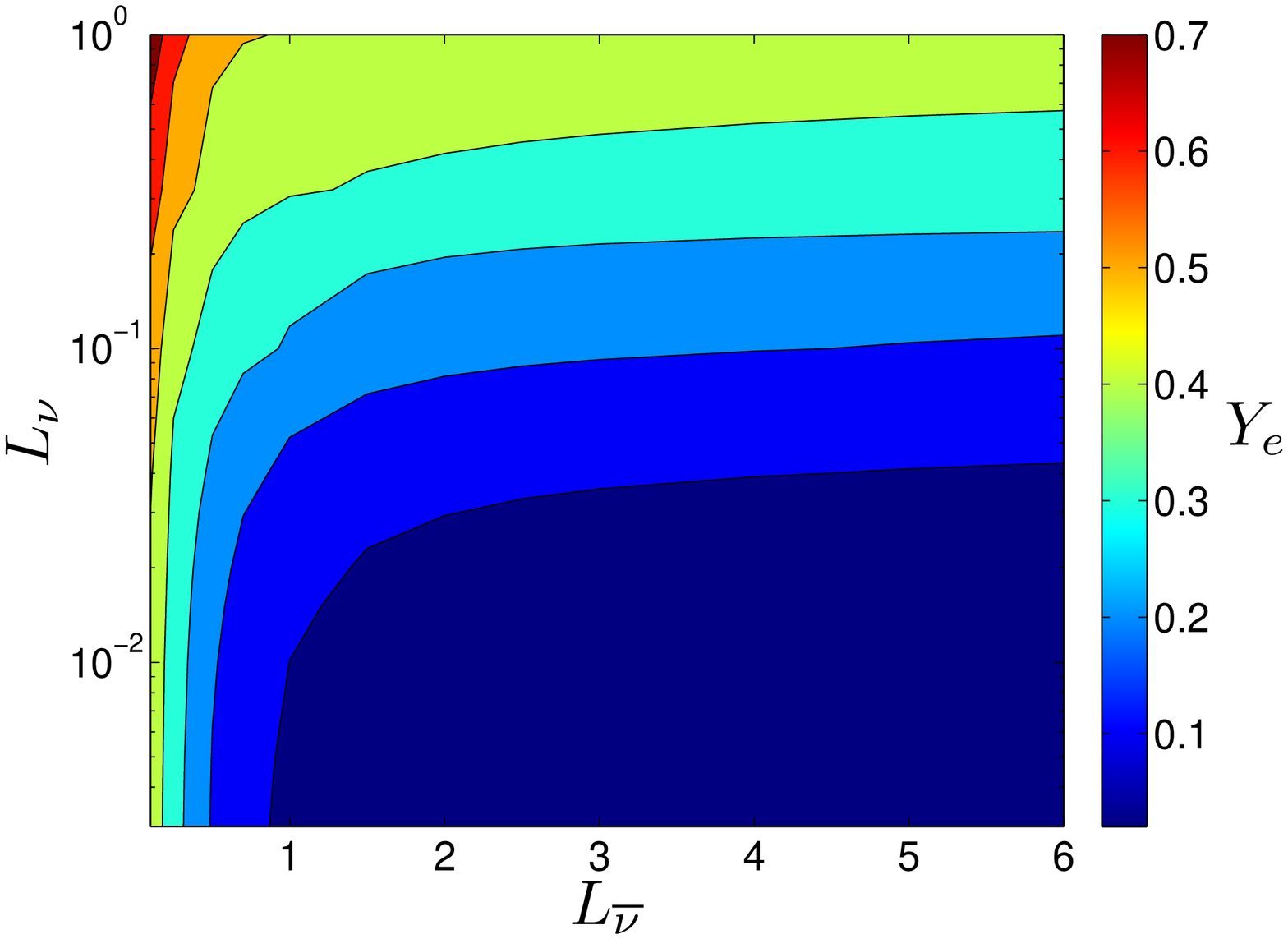}
}
\caption{ The electron fraction, $Y_e$, is shown over a range of electron neutrino and anti-neutrino luminosities,
$L_{\nu}$ and $L_{\overline{\nu}}$. 
A reduction in the electron neutrino luminosity, $L_{\nu}$, prevents the ``alpha'' effect and
leads to a successful $r$-process.
The ``alpha'' effect is avoided as reducing the electron neutrino luminosity, $L_{\nu}$,
lowers the rate of neutrino capture on neutrons forming protons, Eqn. \ref{eq:nu_n}.
The electron fraction, $Y_e$,
is favorable for the $r$-process before $\alpha$-particle formation ($T_9 \approx 9.4$), Fig. \ref{fig:lumyea},
and reductions in $L_{\nu}$ lead to a low $Y_e$ conditions for the $r$-process, Fig. \ref{fig:lumyeb}.
The electron neutrino and anti-neutrino luminosities, $L_{\nu}$ and $L_{\overline{\nu}}$, are in units 
of $\times 10^{51} \ {\rm \ ergs \ s^{-1}}$.
The effective temperature of the electron neutrinos is $T_{\nu} = 3.5 \ {\rm MeV}$ and for the electron anti-neutrinos is
$T_{\overline{\nu}} = 4.5 \ {\rm MeV}$.
For low values of both $L_{\nu}$ and $L_{\overline{\nu}}$, electron and positron capture set the $Y_e$.}
\label{fig:lumye} 
\end{figure}

For reductions in the electron neutrino luminosity relative to the traditional wind model,
but not in the electron anti-neutrino luminosity,
a very large variety of $Y_e$'s result, $0.1 \lesssim Y_e \lesssim 0.8$, as shown in Fig. \ref{fig:lumyea}.
Similar to the case with effective neutrino temperatures, lowering the electron neutrino luminosities, $L_{\nu_e}$, 
decreases the $Y_e$ and leads to neutron-rich conditions.
These early $Y_e$'s from the changed rates in Eqn. \ref{eq:nu_n} and Eqn. \ref{eq:anu_p}
are again consistent with the weak equilibrium values.
For all of Fig. \ref{fig:lumye}, we take the effective electron neutrino temperature to be $T_{\nu_e} = 3.5 {\rm \ MeV}$
and the electron anti-neutrino temperature as $T_{\overline{\nu}_e} = 4.5 {\rm \ MeV}$.

\begin{figure}
\epsscale{1.25}
\plotone{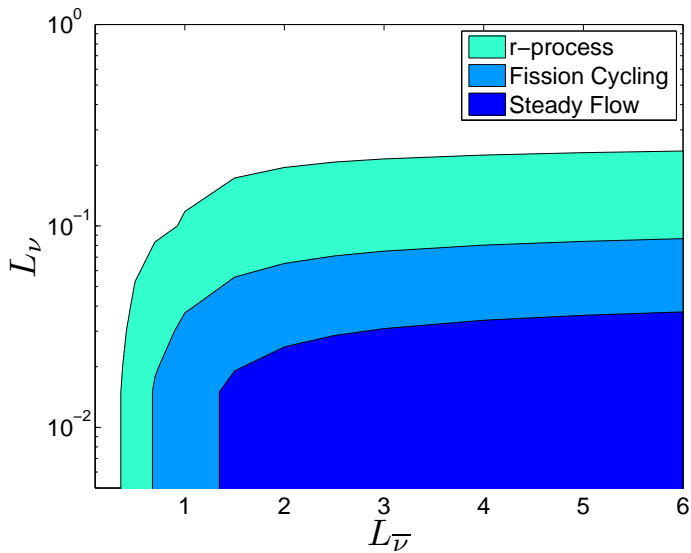}
\caption{ Under the wind conditions of Fig. \ref{fig:lumye}, neutrino luminosities,
$L_{\nu}$ and $L_{\overline{\nu}}$ in units of $\times 10^{51} \ {\rm ergs }$,
necessary for a successful $r$-process (all shaded regions),
the presence of fission cycling (medium and dark blue), 
and for the presence of steady $\beta$-flow (dark blue) are shown.
\label{fig:lum_regions}}
\end{figure}

Near the $r$-process epoch, $T_9 \approx 2.5$, changes to the neutrino luminosities continue to produce a large spread of $Y_e$'s,
as depicted in Fig. \ref{fig:lumyeb}.
A successful $r$-process can occur for reductions to the electron neutrino luminosity of 
$L_{\nu_e} \lesssim 2.3 \times 10^{50} {\rm \ ergs \ s^{-1}}$.
For these decreased electron neutrino luminosities,
the depletion of free neutrons by the alpha effect is diminished 
as fewer new protons are made available through Eqn. \ref{eq:nu_n}.
The neutrino luminosities that lead to a successful $r$-process are shown by all three shaded regions in Fig. \ref{fig:lum_regions}.

We note that neutron-rich conditions beyond the minimum required to produce an $r$-process
occur in our study of the neutrino luminosities.
In fact, order of magnitude reductions of $L_{\nu_e}$ lead to conditions where fission cycling can occur in the $r$-process.
For a $\tau = 0.3 {\rm \ s}$ wind timescale, $Y_e \lesssim 0.17$ are sufficient to lead to fission cycling,
depicted in the medium and dark blue regions of Fig. \ref{fig:lum_regions}.
Conditions even more neutron-rich, $Y_e \lesssim 0.1$,
 lead to a steady flow solution within fission cycling, as described in 
Sec. \ref{steadyflow}, reproducing the general features of the halo star data
and providing an intriguing solution for the realization of a main $r$-process.
This scenario is shown in Fig. \ref{fig:lum_regions} as the dark blue region.
The necessary large reductions to the $L_{\nu_e}$ could be a consequence of 
either active-sterile neutrino oscillations \citep{2006PhRvD..73i3007B} or other new physics.

\section{Conclusions} 
\label{conclusions}

In the neutrino-driven wind, 
fission cycling during $r$-process nucleosynthesis occurs for low electron fractions,
and when combined with steady $\beta$-flow, reproduces the basic features of the main $r$-process.
During fission cycling, material effectively captures out of the third peak and returns to the second peak by fission processes,
linking the second and third peaks, and creating a main $r$-process.
For $0.1 \lesssim Y_e \lesssim 0.3$, these abundance patterns retain some dependence on the initial $Y_e$. 

At even lower electron fractions, $Y_e \lesssim 0.1$, 
steady-$\beta$ flow occurs during the $r$-process 
and the abundance patterns produced are consistent over small changes to the astrophysical conditions.
Steady-$\beta$ flow is a consequence of the presence of 
both fission cycling and $(n,\gamma) \leftrightharpoons (\gamma,n)$ equilibrium.
Fission cycling creates seed nuclei at the second peak,
and $(n,\gamma) \leftrightharpoons (\gamma,n)$ equilibrium determines the $r$-process path.
After $(n,\gamma) \leftrightharpoons (\gamma,n)$ equilibrium fixes the path,
$\beta$-decay between the isotopic chains equilibrate as the first and last chains are connected by fission.
This produces an abundance pattern which is not sensitive to the exact initial $Y_e$. 
The steady-$\beta$ flow abundances strongly depend on the details of fission.

As the final $r$-process abundance pattern is very sensitive to the details of fission, see Fig. \ref{peakyefis},
it is paramount to improve our understanding of which heavy nuclides participate in fission during the $r$-process and
to develop a more precise determination of the daughter products that result.
The presence of fission in the $r$-process may provide a termination point for the $r$-process,
furthering the understanding of which heavy nuclides play a role in the $r$-process.
Under neutron-rich conditions,
the distribution of fission daughter products influences the shape of the $r$-process.
If the $r$-process occurs in very neutron-rich environments,
the fission daughter product distribution also determines the starting isotopic chain of steady-$\beta$ flow,
impacting the region where steady-$\beta$ flow abundances form and
may be important to the understanding of the formation of a main $r$-process.

Low $Y_e$'s at the $r$-process epoch are not found in the traditional neutrino-driven wind.
Before the onset of alpha particle formation, a large range of electron neutrino and anti-neutrino
effective temperatures and luminosities would yield low $Y_e$ environments.
The electron fraction is driven to $Y_e \approx 1/2$ as alpha particle formation binds protons and neutrons in equal numbers.
Changes to the effective temperatures in the ranges of $0.5 {\ \rm MeV} < T_{\nu} < 4 {\ \rm MeV}$ for electron neutrinos
and  $3 {\ \rm MeV} < T_{\overline{\nu}} < 7 {\ \rm MeV}$ for electron anti-neutrinos
result in environments with an ``alpha'' effect and do not yield a main $r$-process.
Given the physics discussed in this paper, realization of the main $r$-process in a conventional neutrino-driven wind is a challenge.  
However, it can occur with a reduction in the electron neutrino capture rate while the electron 
anti-neutrino capture rate remains unchanged.
Future studies on the details of neutrinos, particularly in the post-core bounce supernova environment,
as well as physics beyond the standard model, would benefit the understanding of an $r$-process in the neutrino-driven wind.

\acknowledgments
This work was partially supported by the Department of Energy under contracts DE-FG05-05ER41398 (RS) and DE-FG02-02ER41216 (GCM).
This work was partially supported by the United States National Science Foundation under contract PHY-0244783 (WRH).
Oak Ridge National Laboratory (WRH) is managed by UT-Battelle, LLC, for the U.S. Department of Energy under 
contract DE-AC05-000R22725.  
We acknowledge useful discussions with Gabriel Martinez-Pindedo and Fredrich-Karl Thielemann.


%
%
%
%


\makeatletter
\let\jnl@style=\rm
\def\ref@jnl#1{{\jnl@style#1}}
\def\ref@jnl#1{{#1}}
\def\ref@jnl{{\jnl@style}}

\def\aj{\ref@jnl{AJ}}                   
\def\araa{\ref@jnl{ARA\&A}}             
\def\apj{\ref@jnl{ApJ}}                 
\def\apjl{\ref@jnl{ApJ}}                
\def\apjs{\ref@jnl{ApJS}}               
\def\ao{\ref@jnl{Appl.~Opt.}}           
\def\apss{\ref@jnl{Ap\&SS}}             
\def\aap{\ref@jnl{A\&A}}                
\def\aapr{\ref@jnl{A\&A~Rev.}}          
\def\aaps{\ref@jnl{A\&AS}}              
\def\azh{\ref@jnl{AZh}}                 
\def\baas{\ref@jnl{BAAS}}               
\def\jrasc{\ref@jnl{JRASC}}             
\def\memras{\ref@jnl{MmRAS}}            
\def\mnras{\ref@jnl{MNRAS}}             
\def\pra{\ref@jnl{Phys.~Rev.~A}}        
\def\prb{\ref@jnl{Phys.~Rev.~B}}        
\def\prc{\ref@jnl{Phys.~Rev.~C}}        
\def\prd{\ref@jnl{Phys.~Rev.~D}}        
\def\pre{\ref@jnl{Phys.~Rev.~E}}        
\def\prl{\ref@jnl{Phys.~Rev.~Lett.}}    
\def\pasp{\ref@jnl{PASP}}               
\def\pasj{\ref@jnl{PASJ}}               
\def\qjras{\ref@jnl{QJRAS}}             
\def\skytel{\ref@jnl{S\&T}}             
\def\solphys{\ref@jnl{Sol.~Phys.}}      
\def\sovast{\ref@jnl{Soviet~Ast.}}      
\def\ssr{\ref@jnl{Space~Sci.~Rev.}}     
\def\zap{\ref@jnl{ZAp}}                 
\def\nat{\ref@jnl{Nature}}              
\def\iaucirc{\ref@jnl{IAU~Circ.}}       
\def\aplett{\ref@jnl{Astrophys.~Lett.}} 
\def\apspr{\ref@jnl{Astrophys.~Space~Phys.~Res.}}
\def\bain{\ref@jnl{Bull.~Astron.~Inst.~Netherlands}} 
\def\fcp{\ref@jnl{Fund.~Cosmic~Phys.}}  
\def\gca{\ref@jnl{Geochim.~Cosmochim.~Acta}}   
\def\grl{\ref@jnl{Geophys.~Res.~Lett.}} 
\def\jcp{\ref@jnl{J.~Chem.~Phys.}}      
\def\jgr{\ref@jnl{J.~Geophys.~Res.}}    
\def\jqsrt{\ref@jnl{J.~Quant.~Spec.~Radiat.~Transf.}}
\def\memsai{\ref@jnl{Mem.~Soc.~Astron.~Italiana}}
\def\nphysa{\ref@jnl{Nucl.~Phys.~A}}   
\def\physrep{\ref@jnl{Phys.~Rep.}}   
\def\physscr{\ref@jnl{Phys.~Scr}}   
\def\planss{\ref@jnl{Planet.~Space~Sci.}}   
\def\procspie{\ref@jnl{Proc.~SPIE}}   

\let\astap=\aap
\let\apjlett=\apjl
\let\apjsupp=\apjs
\let\applopt=\ao





\bibliographystyle{apj}

\begin{thebibliography}{73}
\expandafter\ifx\csname natexlab\endcsname\relax\def\natexlab#1{#1}\fi

\bibitem[{Arcones {et~al.}(2006)Arcones, Janka, \& Scheck}]{Arcones:2006uq}
Arcones, A., Janka, H.-T., \& Scheck, L. 2006

\bibitem[{Argast {et~al.}(2004)Argast, Samland, Thielemann, \&
  Qian}]{argast-2004-416}
Argast, D., Samland, M., Thielemann, F.~K., \& Qian, Y.~Z. 2004,
  ASTRON.ASTROPHYS., 416, 997

\bibitem[{{Arnould} {et~al.}(2007){Arnould}, {Goriely}, \&
  {Takahashi}}]{2007arXiv0705.4512A}
{Arnould}, M., {Goriely}, S., \& {Takahashi}, K. 2007, ArXiv e-prints, 705

\bibitem[{{Beun} {et~al.}(2006){Beun}, {McLaughlin}, {Surman}, \&
  {Hix}}]{2006PhRvD..73i3007B}
{Beun}, J., {McLaughlin}, G.~C., {Surman}, R., \& {Hix}, W.~R. 2006, \prd, 73,
  093007

\bibitem[{Beun {et~al.}(2006)Beun, McLaughlin, Surman, \& Hix}]{Beun:2006vg}
Beun, J., McLaughlin, G.~C., Surman, R., \& Hix, W.~R. 2006, PoS, NIC-IX, 140

\bibitem[{{Bruenn}(1989{\natexlab{a}})}]{1989ApJ...340..955B}
{Bruenn}, S.~W. 1989{\natexlab{a}}, \apj, 340, 955

\bibitem[{{Bruenn}(1989{\natexlab{b}})}]{1989ApJ...341..385B}
---. 1989{\natexlab{b}}, \apj, 341, 385

\bibitem[{{Burbidge} {et~al.}(1957){Burbidge}, {Burbidge}, {Fowler}, \&
  {Hoyle}}]{1957RvMP...29..547B}
{Burbidge}, E.~M., {Burbidge}, G.~R., {Fowler}, W.~A., \& {Hoyle}, F. 1957,
  Reviews of Modern Physics, 29, 547

\bibitem[{{Burris} {et~al.}(2000){Burris}, {Pilachowski}, {Armandroff},
  {Sneden}, {Cowan}, \& {Roe}}]{2000ApJ...544..302B}
{Burris}, D.~L., {Pilachowski}, C.~A., {Armandroff}, T.~E., {Sneden}, C.,
  {Cowan}, J.~J., \& {Roe}, H. 2000, \apj, 544, 302

\bibitem[{{Burrows} {et~al.}(2007){Burrows}, {Dessart}, {Livne}, {Ott}, \&
  {Murphy}}]{2007astro.ph..2539B}
{Burrows}, A., {Dessart}, L., {Livne}, E., {Ott}, C.~D., \& {Murphy}, J. 2007,
  ArXiv Astrophysics e-prints

\bibitem[{{Cameron}(1957)}]{1957Cameron}
{Cameron}, A.~G.~W. 1957, Chalk River Rep.

\bibitem[{{Cardall} \& {Fuller}(1997)}]{1997ApJ...486L.111C}
{Cardall}, C.~Y. \& {Fuller}, G.~M. 1997, \apjl, 486, L111+

\bibitem[{{Cayrel} {et~al.}(2001){Cayrel}, {Hill}, {Beers}, {Barbuy}, {Spite},
  {Spite}, {Plez}, {Andersen}, {Bonifacio}, {Fran{\c c}ois}, {Molaro},
  {Nordstr{\"o}m}, \& {Primas}}]{2001Natur.409..691C}
{Cayrel}, R., {Hill}, V., {Beers}, T.~C., {Barbuy}, B., {Spite}, M., {Spite},
  F., {Plez}, B., {Andersen}, J., {Bonifacio}, P., {Fran{\c c}ois}, P.,
  {Molaro}, P., {Nordstr{\"o}m}, B., \& {Primas}, F. 2001, \nat, 409, 691

\bibitem[{{Christlieb} {et~al.}(2004){Christlieb}, {Beers}, {Barklem},
  {Bessell}, {Hill}, {Holmberg}, {Korn}, {Marsteller}, {Mashonkina}, {Qian},
  {Rossi}, {Wasserburg}, {Zickgraf}, {Kratz}, {Nordstr{\"o}m}, {Pfeiffer},
  {Rhee}, \& {Ryan}}]{2004A&A...428.1027C}
{Christlieb}, N., {Beers}, T.~C., {Barklem}, P.~S., {Bessell}, M., {Hill}, V.,
  {Holmberg}, J., {Korn}, A.~J., {Marsteller}, B., {Mashonkina}, L., {Qian},
  Y.-Z., {Rossi}, S., {Wasserburg}, G.~J., {Zickgraf}, F.-J., {Kratz}, K.-L.,
  {Nordstr{\"o}m}, B., {Pfeiffer}, B., {Rhee}, J., \& {Ryan}, S.~G. 2004, \aap,
  428, 1027

\bibitem[{{Cowan} {et~al.}(1999){Cowan}, {Pfeiffer}, {Kratz}, {Thielemann},
  {Sneden}, {Burles}, {Tytler}, \& {Beers}}]{1999ApJ...521..194C}
{Cowan}, J.~J., {Pfeiffer}, B., {Kratz}, K.-L., {Thielemann}, F.-K., {Sneden},
  C., {Burles}, S., {Tytler}, D., \& {Beers}, T.~C. 1999, \apj, 521, 194

\bibitem[{{Cowan} {et~al.}(2002){Cowan}, {Sneden}, {Burles}, {Ivans}, {Beers},
  {Truran}, {Lawler}, {Primas}, {Fuller}, {Pfeiffer}, \&
  {Kratz}}]{2002ApJ...572..861C}
{Cowan}, J.~J., {Sneden}, C., {Burles}, S., {Ivans}, I.~I., {Beers}, T.~C.,
  {Truran}, J.~W., {Lawler}, J.~E., {Primas}, F., {Fuller}, G.~M., {Pfeiffer},
  B., \& {Kratz}, K.-L. 2002, \apj, 572, 861

\bibitem[{Cowan {et~al.}(2006)Cowan, Sneden, Lawler, \&
  Hartog}]{2006astro.ph.10412C}
Cowan, J.~J., Sneden, C., Lawler, J.~E., \& Hartog, E. A.~D. 2006, PoS, NIC-IX,
  014

\bibitem[{{Cowan} {et~al.}(1987){Cowan}, {Thielemann}, \&
  {Truran}}]{1987ApJ...323..543C}
{Cowan}, J.~J., {Thielemann}, F.-K., \& {Truran}, J.~W. 1987, \apj, 323, 543

\bibitem[{{Cowan} {et~al.}(1991){Cowan}, {Thielemann}, \&
  {Truran}}]{1991PhR...208..267C}
---. 1991, \physrep, 208, 267

\bibitem[{Duflo \& Zuker(1995)}]{PhysRevC.52.R23}
Duflo, J. \& Zuker, A. 1995, Phys. Rev. C, 52, R23

\bibitem[{{Duncan} {et~al.}(1986){Duncan}, {Shapiro}, \&
  {Wasserman}}]{1986ApJ...309..141D}
{Duncan}, R.~C., {Shapiro}, S.~L., \& {Wasserman}, I. 1986, \apj, 309, 141

\bibitem[{{Fetter} {et~al.}(2003){Fetter}, {McLaughlin}, {Balantekin}, \&
  {Fuller}}]{2003APh....18..433F}
{Fetter}, J., {McLaughlin}, G.~C., {Balantekin}, A.~B., \& {Fuller}, G.~M.
  2003, Astroparticle Physics, 18, 433

\bibitem[{{Frebel} {et~al.}(2007){Frebel}, {Christlieb}, {Norris}, {Thom},
  {Beers}, \& {Rhee}}]{2007ApJ...660L.117F}
{Frebel}, A., {Christlieb}, N., {Norris}, J.~E., {Thom}, C., {Beers}, T.~C., \&
  {Rhee}, J. 2007, \apjl, 660, L117

\bibitem[{{Freiburghaus} {et~al.}(1999){Freiburghaus}, {Rembges}, {Rauscher},
  {Kolbe}, {Thielemann}, {Kratz}, {Pfeiffer}, \& {Cowan}}]{1999ApJ...516..381F}
{Freiburghaus}, C., {Rembges}, J.-F., {Rauscher}, T., {Kolbe}, E.,
  {Thielemann}, F.-K., {Kratz}, K.-L., {Pfeiffer}, B., \& {Cowan}, J.~J. 1999,
  \apj, 516, 381

\bibitem[{{Fryer} {et~al.}(2006){Fryer}, {Herwig}, {Hungerford}, \&
  {Timmes}}]{2006ApJ...646L.131F}
{Fryer}, C.~L., {Herwig}, F., {Hungerford}, A., \& {Timmes}, F.~X. 2006, \apjl,
  646, L131

\bibitem[{{Fuller} \& {Meyer}(1995)}]{1995ApJ...453..792F}
{Fuller}, G.~M. \& {Meyer}, B.~S. 1995, \apj, 453, 792

\bibitem[{{Goriely} {et~al.}(2005){Goriely}, {Demetriou}, {Janka}, {Pearson},
  \& {Samyn}}]{2005NuPhA.758..587G}
{Goriely}, S., {Demetriou}, P., {Janka}, H.-T., {Pearson}, J.~M., \& {Samyn},
  M. 2005, Nuclear Physics A, 758, 587

\bibitem[{{Hilf} {et~al.}(1976)}]{hilf76}
{Hilf}, E.~R. {et~al.} 1976, Suppl. to Proc. Int. Conf. NFFS-3, 76-13, 142

\bibitem[{{Hill} {et~al.}(2002){Hill}, {Plez}, {Cayrel}, {Beers},
  {Nordstr{\"o}m}, {Andersen}, {Spite}, {Spite}, {Barbuy}, {Bonifacio},
  {Depagne}, {Fran{\c c}ois}, \& {Primas}}]{2002A&A...387..560H}
{Hill}, V., {Plez}, B., {Cayrel}, R., {Beers}, T.~C., {Nordstr{\"o}m}, B.,
  {Andersen}, J., {Spite}, M., {Spite}, F., {Barbuy}, B., {Bonifacio}, P.,
  {Depagne}, E., {Fran{\c c}ois}, P., \& {Primas}, F. 2002, \aap, 387, 560

\bibitem[{Hix \& Thielemann(1999)}]{hix1999}
Hix, W.~R. \& Thielemann, F.~K. 1999, J. Comp. App. Math., 109, 321

\bibitem[{{Hoffman} {et~al.}(1997){Hoffman}, {Woosley}, \&
  {Qian}}]{1997ApJ...482..951H}
{Hoffman}, R.~D., {Woosley}, S.~E., \& {Qian}, Y.-Z. 1997, \apj, 482, 951

\bibitem[{{Honda} {et~al.}(2004){Honda}, {Aoki}, {Kajino}, {Ando}, {Beers},
  {Izumiura}, {Sadakane}, \& {Takada-Hidai}}]{2004ApJ...607..474H}
{Honda}, S., {Aoki}, W., {Kajino}, T., {Ando}, H., {Beers}, T.~C., {Izumiura},
  H., {Sadakane}, K., \& {Takada-Hidai}, M. 2004, \apj, 607, 474

\bibitem[{{Ivans} {et~al.}(2006){Ivans}, {Simmerer}, {Sneden}, {Lawler},
  {Cowan}, {Gallino}, \& {Bisterzo}}]{2006ApJ...645..613I}
{Ivans}, I.~I., {Simmerer}, J., {Sneden}, C., {Lawler}, J.~E., {Cowan}, J.~J.,
  {Gallino}, R., \& {Bisterzo}, S. 2006, \apj, 645, 613

\bibitem[{{Keli{\'c}} {et~al.}(2005){Keli{\'c}}, {Zinner}, {Kolbe}, {Langanke},
  \& {Schmidt}}]{2005PhLB..616...48K}
{Keli{\'c}}, A., {Zinner}, N., {Kolbe}, E., {Langanke}, K., \& {Schmidt}, K.-H.
  2005, Physics Letters B, 616, 48

\bibitem[{{Kolbe} {et~al.}(2004){Kolbe}, {Langanke}, \&
  {Fuller}}]{2004PhRvL..92k1101K}
{Kolbe}, E., {Langanke}, K., \& {Fuller}, G.~M. 2004, Physical Review Letters,
  92, 111101

\bibitem[{Martinez-Pinedo {et~al.}(2006)}]{MartinezPinedo:2006fk}
Martinez-Pinedo, G. {et~al.} 2006, PoS, NIC-IX, 064

\bibitem[{{McLaughlin} {et~al.}(1999){McLaughlin}, {Fetter}, {Balantekin}, \&
  {Fuller}}]{1999PhRvC..59.2873M}
{McLaughlin}, G.~C., {Fetter}, J.~M., {Balantekin}, A.~B., \& {Fuller}, G.~M.
  1999, \prc, 59, 2873

\bibitem[{{McLaughlin} \& {Fuller}(1995)}]{1995ApJ...455..202M}
{McLaughlin}, G.~C. \& {Fuller}, G.~M. 1995, \apj, 455, 202

\bibitem[{{McLaughlin} {et~al.}(1996){McLaughlin}, {Fuller}, \&
  {Wilson}}]{1996ApJ...472..440M}
{McLaughlin}, G.~C., {Fuller}, G.~M., \& {Wilson}, J.~R. 1996, \apj, 472, 440

\bibitem[{{McWilliam}(1998)}]{1998AJ....115.1640M}
{McWilliam}, A. 1998, \aj, 115, 1640

\bibitem[{{McWilliam} {et~al.}(1995){McWilliam}, {Preston}, {Sneden}, \&
  {Searle}}]{1995AJ....109.2757M}
{McWilliam}, A., {Preston}, G.~W., {Sneden}, C., \& {Searle}, L. 1995, \aj,
  109, 2757

\bibitem[{{Metzger} {et~al.}(2007){Metzger}, {Thompson}, \&
  {Quataert}}]{2007ApJ...659..561M}
{Metzger}, B.~D., {Thompson}, T.~A., \& {Quataert}, E. 2007, \apj, 659, 561

\bibitem[{{Meyer}(1989)}]{1989ApJ...343..254M}
{Meyer}, B.~S. 1989, \apj, 343, 254

\bibitem[{{Meyer}(1994)}]{1994ARA&A..32..153M}
---. 1994, \araa, 32, 153

\bibitem[{{Meyer} {et~al.}(1992){Meyer}, {Mathews}, {Howard}, {Woosley}, \&
  {Hoffman}}]{1992ApJ...399..656M}
{Meyer}, B.~S., {Mathews}, G.~J., {Howard}, W.~M., {Woosley}, S.~E., \&
  {Hoffman}, R.~D. 1992, \apj, 399, 656

\bibitem[{{Meyer} {et~al.}(1998){Meyer}, {McLaughlin}, \&
  {Fuller}}]{1998PhRvC..58.3696M}
{Meyer}, B.~S., {McLaughlin}, G.~C., \& {Fuller}, G.~M. 1998, \prc, 58, 3696

\bibitem[{{M{\"o}ller} {et~al.}(1997){M{\"o}ller}, {Nix}, \&
  {Kratz}}]{1997ADNDT..66..131M}
{M{\"o}ller}, P., {Nix}, J.~R., \& {Kratz}, K.-L. 1997, Atomic Data and Nuclear
  Data Tables, 66, 131

\bibitem[{{Otsuki} {et~al.}(2000){Otsuki}, {Tagoshi}, {Kajino}, \&
  {Wanajo}}]{2000ApJ...533..424O}
{Otsuki}, K., {Tagoshi}, H., {Kajino}, T., \& {Wanajo}, S.-y. 2000, \apj, 533,
  424

\bibitem[{{Panov} {et~al.}(2005){Panov}, {Kolbe}, {Pfeiffer}, {Rauscher},
  {Kratz}, \& {Thielemann}}]{2005NuPhA.747..633P}
{Panov}, I.~V., {Kolbe}, E., {Pfeiffer}, B., {Rauscher}, T., {Kratz}, K.-L., \&
  {Thielemann}, F.-K. 2005, Nuclear Physics A, 747, 633

\bibitem[{{Panov} \& {Thielemann}(2004)}]{2004AstL...30..647P}
{Panov}, I.~V. \& {Thielemann}, F.-K. 2004, Astronomy Letters, 30, 647

\bibitem[{{Pearson} {et~al.}(1996){Pearson}, {Nayak}, \&
  {Goriely}}]{1996PhLB..387..455P}
{Pearson}, J.~M., {Nayak}, R.~C., \& {Goriely}, S. 1996, Physics Letters B,
  387, 455

\bibitem[{{Qian}(2003)}]{2003PrPNP..50..153Q}
{Qian}, Y.-Z. 2003, Progress in Particle and Nuclear Physics, 50, 153

\bibitem[{{Qian} \& {Woosley}(1996)}]{1996ApJ...471..331Q}
{Qian}, Y.-Z. \& {Woosley}, S.~E. 1996, \apj, 471, 331

\bibitem[{{Rauscher} {et~al.}(1994){Rauscher}, {Applegate}, {Cowan},
  {Thielemann}, \& {Wiescher}}]{1994ApJ...429..499R}
{Rauscher}, T., {Applegate}, J.~H., {Cowan}, J.~J., {Thielemann}, F.-K., \&
  {Wiescher}, M. 1994, \apj, 429, 499

\bibitem[{{Rauscher} \& {Thielemann}(2000)}]{2000ADNDT..75....1R}
{Rauscher}, T. \& {Thielemann}, F.-K. 2000, Atomic Data and Nuclear Data
  Tables, 75, 1

\bibitem[{{Rosswog} {et~al.}(2003){Rosswog}, {Ramirez-Ruiz}, \&
  {Davies}}]{2003MNRAS.345.1077R}
{Rosswog}, S., {Ramirez-Ruiz}, E., \& {Davies}, M.~B. 2003, \mnras, 345, 1077

\bibitem[{{Seeger} {et~al.}(1965){Seeger}, {Fowler}, \&
  {Clayton}}]{1965ApJS...11..121S}
{Seeger}, P.~A., {Fowler}, W.~A., \& {Clayton}, D.~D. 1965, \apjs, 11, 121

\bibitem[{Simmerer {et~al.}(2004)}]{Simmerer:2004jq}
Simmerer, J. {et~al.} 2004, Astrophys. J., 617, 1091

\bibitem[{{Sneden} {et~al.}(2003){Sneden}, {Cowan}, {Lawler}, {Ivans},
  {Burles}, {Beers}, {Primas}, {Hill}, {Truran}, {Fuller}, {Pfeiffer}, \&
  {Kratz}}]{2003ApJ...591..936S}
{Sneden}, C., {Cowan}, J.~J., {Lawler}, J.~E., {Ivans}, I.~I., {Burles}, S.,
  {Beers}, T.~C., {Primas}, F., {Hill}, V., {Truran}, J.~W., {Fuller}, G.~M.,
  {Pfeiffer}, B., \& {Kratz}, K.-L. 2003, \apj, 591, 936

\bibitem[{{Sneden} {et~al.}(1996){Sneden}, {McWilliam}, {Preston}, {Cowan},
  {Burris}, \& {Armosky}}]{1996ApJ...467..819S}
{Sneden}, C., {McWilliam}, A., {Preston}, G.~W., {Cowan}, J.~J., {Burris},
  D.~L., \& {Armosky}, B.~J. 1996, \apj, 467, 819

\bibitem[{{Sumiyoshi} {et~al.}(2001){Sumiyoshi}, {Terasawa}, {Mathews},
  {Kajino}, {Yamada}, \& {Suzuki}}]{2001ApJ...562..880S}
{Sumiyoshi}, K., {Terasawa}, M., {Mathews}, G.~J., {Kajino}, T., {Yamada}, S.,
  \& {Suzuki}, H. 2001, \apj, 562, 880

\bibitem[{{Surman} \& {Engel}(2001)}]{2001PhRvC..64c5801S}
{Surman}, R. \& {Engel}, J. 2001, \prc, 64, 035801

\bibitem[{Surman {et~al.}(1997)Surman, Engel, Bennett, \&
  Meyer}]{PhysRevLett.79.1809}
Surman, R., Engel, J., Bennett, J.~R., \& Meyer, B.~S. 1997, Phys. Rev. Lett.,
  79, 1809

\bibitem[{{Surman} \& {McLaughlin}(2004)}]{2004ApJ...603..611S}
{Surman}, R. \& {McLaughlin}, G.~C. 2004, \apj, 603, 611

\bibitem[{Surman {et~al.}(2006)Surman, McLaughlin, \& Hix}]{Surman:2005kf}
Surman, R., McLaughlin, G.~C., \& Hix, W.~R. 2006, Astrophys. J., 643, 1057

\bibitem[{{Terasawa} {et~al.}(2004){Terasawa}, {Langanke}, {Kajino}, {Mathews},
  \& {Kolbe}}]{2004ApJ...608..470T}
{Terasawa}, M., {Langanke}, K., {Kajino}, T., {Mathews}, G.~J., \& {Kolbe}, E.
  2004, \apj, 608, 470

\bibitem[{Thielemann {et~al.}(2001)}]{Thielemann:2001rn}
Thielemann, F.~K. {et~al.} 2001, Prog. Part. Nucl. Phys., 46, 5

\bibitem[{{Thompson} {et~al.}(2001){Thompson}, {Burrows}, \&
  {Meyer}}]{2001ApJ...562..887T}
{Thompson}, T.~A., {Burrows}, A., \& {Meyer}, B.~S. 2001, \apj, 562, 887

\bibitem[{{Wanajo}(2007)}]{2007arXiv0706.4360W}
{Wanajo}, S. 2007, ArXiv e-prints, 706

\bibitem[{{Wanajo} {et~al.}(2001){Wanajo}, {Kajino}, {Mathews}, \&
  {Otsuki}}]{2001ApJ...554..578W}
{Wanajo}, S., {Kajino}, T., {Mathews}, G.~J., \& {Otsuki}, K. 2001, \apj, 554,
  578

\bibitem[{{Wanajo} {et~al.}(2003){Wanajo}, {Tamamura}, {Itoh}, {Nomoto},
  {Ishimaru}, {Beers}, \& {Nozawa}}]{2003ApJ...593..968W}
{Wanajo}, S., {Tamamura}, M., {Itoh}, N., {Nomoto}, K., {Ishimaru}, Y.,
  {Beers}, T.~C., \& {Nozawa}, S. 2003, \apj, 593, 968

\bibitem[{{Westin} {et~al.}(2000){Westin}, {Sneden}, {Gustafsson}, \&
  {Cowan}}]{2000ApJ...530..783W}
{Westin}, J., {Sneden}, C., {Gustafsson}, B., \& {Cowan}, J.~J. 2000, \apj,
  530, 783

\bibitem[{{Woosley} {et~al.}(1994){Woosley}, {Wilson}, {Mathews}, {Hoffman}, \&
  {Meyer}}]{1994ApJ...433..229W}
{Woosley}, S.~E., {Wilson}, J.~R., {Mathews}, G.~J., {Hoffman}, R.~D., \&
  {Meyer}, B.~S. 1994, \apj, 433, 229

\end{thebibliography}



\end{document}